\documentclass[a, pdf]{iucr}              

\usepackage{color,xcolor}
\usepackage{xspace}
\RequirePackage{graphicx}
\usepackage{epstopdf}
\usepackage{mathtools, amsmath, amsthm, amssymb, amsfonts}
\usepackage[12pt]{moresize}
\usepackage{multirow}
\usepackage{enumitem}
\usepackage{url}

\usepackage{enumitem}

\definecolor{timgreen}{HTML}{106d00}

\definecolor{sborange}{HTML}{ce6e00}

\definecolor{citationtags}{HTML}{21a889}

\definecolor{finedits}{HTML}{b70c00}

\definecolor{dgreen}{HTML}{008000}

\newcommand{\be}{\begin{enumerate}[wide, labelwidth=!, labelindent=0pt,
label=\textbf{\textcolor{purple}{\arabic*}.}]}
\newcommand{\bei}{\begin{enumerate}}

\newcommand{\ee}{\end{enumerate}}

\newcounter{saveenumi}

\newcommand{\est}{$\sim$}
\newcommand{\npc}{Fig.~\ref{fig:npCore}}
\newcommand{\pdfcc}{Fig.~\ref{fig:PdFCC}}
\newcommand{\pdtwin}{Fig.~\ref{fig:PdTwin}}
\newcommand{\scsix}{Fig.~\ref{fig:SC6}}
\newcommand{\poly}{Fig.~\ref{fig:polymorphism}}

\newcommand{\rw}{R$_w$}

\newcommand{\pdfgui}{\textsc{PDFgui}\xspace}
\newcommand{\diffpy}{\textsc{diffpy}\xspace}
\newcommand{\pdfgetxthree}{\textsc{PDFgetX3}\xspace}

\newcommand{\fittwod}{\textsc{Fit2D}\xspace}

\newcommand{\xpdf}{\textsc{xPDFsuite}\xspace}
\newcommand{\cmi}{\textsc{CMI}\xspace}

\newcommand{\ase}{\textsc{ASE}\xspace}
\newcommand{\dpc}{\textsc{DebyePDFCalculator}\xspace}

\newcommand{\srfit}{\textsc{SrFit}\xspace}
\newcommand{\cmining}{\textsc{cluster-mining}\xspace}

\newcommand{\minemap}{cluster-screen map\xspace}

\journalcode{A}              

\begin{document}                  



\title{Cluster-mining: An approach for determining core structures of metallic nanoparticles
from atomic pair distribution function data}

\shorttitle{cluster-mining}

\author[a]{Soham}{Banerjee}
\author[a]{Chia-Hao}{Liu}
\author[d]{Kirsten~M.~\O.}{Jensen}
\author[b]{Pavol}{Juh\'{a}s}
\author[e]{Jennifer~D.}{Lee}
\author[h]{Marcus}{Tofanelli}
\author[g]{Christopher~J.}{Ackerson}
\author[e,f]{Christopher~B.}{Murray}
\cauthor[a,c]{Simon~J.~L.}{Billinge}{sb2896@columbia.edu}

\aff[a]{{Department of Applied Physics and Applied Mathematics, Columbia University},
\city{{New York}, New York, 10027, \country{USA}}}

\aff[b]{{Computational Science Initiative, Brookhaven National Laboratory},
\city{Upton, NY~11973}, \country{USA}}	

\aff[c]{{Condensed Matter Physics and Materials Science Department, Brookhaven National Laboratory},
\city{Upton, New York, 11973, \country{USA}}}

\aff[d]{{Department of Chemistry, University of Copenhagen},
\city{Copenhagen, DK-2100}, \country{Denmark}}	

\aff[e]{{Department of Chemistry, University of Pennsylvania},
\city{Philadelphia, PA, 19104}, \country{USA}}	

\aff[f]{{Department of Materials Science and Engineering, University of Pennsylvania},
\city{Philadelphia, PA, 19104}, \country{USA}}

\aff[g]{{Department of Chemistry, Colorado State University},
\city{Fort Collins, CO, 80523}, \country{USA}}

\aff[h]{{Department of Chemistry, University of Pittsburgh},
\city{Pittsburgh, PA, 15260}, \country{USA}}









\maketitle
%

\begin{abstract}
We present a novel approach for finding and evaluating structural models of small metallic nanoparticles. Rather than fitting a single model with many degrees of freedom, the approach algorithmically builds libraries of nanoparticle clusters from multiple structural motifs, and individually fits them to experimental PDFs. Each cluster-fit is highly constrained. The approach, called cluster-mining, returns all candidate structure models that are consistent with the data as measured by a goodness of fit.  It is highly automated, easy to use, and yields models that are more physically realistic and result in better agreement to the data than models based on cubic close-packed crystallographic cores, often reported in the literature for metallic nanoparticles.
%

\vspace{3em}
\end{abstract}


\section{Introduction}

Advances in the synthesis of metallic nanoparticles (NP) have given researchers a great deal of control in tailoring their functionalities for many applications including catalysis~\cite{LewisChemCatColloidCluster1993,SomorjaiMolecularFactorsCatalytic2008}, plasmonics~\cite{AtwaterPlasmonicsimprovedphotovoltaic2010, LinicPlasmonicmetalnanostructuresefficient2011}, energy conversion~\cite{AricoNanostructuredmaterialsadvanced2005b}, and biomedicine~\cite{RosiNanostructuresBiodiagnostics2005, AckersonRigidSpecificDiscrete2006, NuneNanoparticlesbiomedicalimaging2009}.
At the simplest level, the distinct properties of nanoparticles can be attributed to the increased role of their external surfaces, which can be manipulated by changing experimental parameters in a synthesis to obtain particles of a certain size, shape and composition.
However, an atomic scale understanding of the structural mechanisms which influence the growth of metallic nanoparticles, resulting in ensembles with varying degrees of uniformity in size, morphology and chemical ordering, remains a major challenge~\cite{Bojesenchemistrynucleation2016a, LeeNonclassicalnucleationgrowth2016}.
Critical to engineering the next generation of these materials by design, rather than empirical optimization, is to develop structural probes and modeling methodologies capable of quantifying the arrangements of atoms at the smallest length-scales possible.

Determining the atomic core structures of ultrasmall nanoparticles using x-ray powder diffraction methods is difficult~\cite{billi;s07}.
The information obtained in these experiments is degraded not only due to finite size effects, but also because the internal arrangements of atoms deviate significantly from bulk materials.
Non-crystallographic structures have long been reported in electron microscopic studies of metallic nanoparticles~\cite{
InoEpitaxialGrowthMetals1966,
InoStabilityMultiplyTwinnedParticles1969,
MarksMultiplytwinnedparticlessilver1979,
SunShapeControlledSynthesisGold2002,
ChenThreedimensionalimagingdislocations2013a} and it is established that growth mechanisms across a diversity of synthesis methods are directed by the size-dependent formation and rearrangement of multiply-twinned domains, in addition to thermodynamic stabilization of nanoparticle surfaces by capping agents~\cite{
LoftonMechanismsControllingCrystal2005,
LangilleStepwiseEvolutionSpherical2012}.
Despite this evidence, atomic models built from fcc cores, which do not account for the multi-domain nature of these materials, are still commonly used in PDF analysis of metallic nanostructures~\cite{petko;prb10,page;jac11i,kumar;jacs14,
FleuryGoldNanoparticleInternal2015a,
WuCompositionStructureActivity2015a,
PoulainSizeAuNanoparticlesSupported2016,
PetkovApplicationdifferentialresonant2018}.

It was recently demonstrated that the atomic pair distribution function (PDF) does contain information allowing for the detection and characterization of internal atomic interfaces in a diversity of metallic nanomaterials and atomic clusters~\cite{BanerjeeImprovedmodelsmetallic2018}.
It was also shown that the PDF could differentiate between various arrangements of multiply-twinned domains.
For a majority of the samples surveyed,  simple decahedral or icosahedral cluster cores, instead of fcc attenuated crystal (AC) approximations or single crystal fcc cutouts, gave significantly improved fits.
This analysis hinged on time-consuming, manual trial-and-error refinements of a few representative cluster models from different structure motifs.
Here we describe a new approach for determining the best models for metallic nanoparticle core structures, by automatically generating large numbers of candidate cluster structures and comparing them to PDF data from nanoparticles.
The methodology differs from traditional approaches for crystallographic analysis of  nanoparticles where a single model containing many refinable parameters is used to fit peak profiles from a measured diffraction pattern.
Instead, this approach uses many structure models and highly constrained refinements to screen libraries of discrete clusters against experimental PDF data, with the aim of finding the most representative cluster structures for the ensemble average nanoparticle from any given synthesis.

\section{Modeling}
\label{sec:modeling}

The core of the new approach is to generate large numbers of candidate structure models, which in principle could be pulled from databases, or generated algorithmically.
PDFs are then computed from each model and compared to a given measured PDF.
A small number of refinable parameters may be varied in this last comparison step, such as an overall scale factor and an average bond-length, in such a way as to minimize an agreement factor, \rw, described in greater detail below.
The results of the comparisons for all models are then reported back to the experimenter.
Our target in the initial implementation described here is to consider different finite-sized cluster models to compare against data collected from small metallic nanoparticle samples and in this case we generate the libraries of clusters, which we call cluster-mines, algorithmically.

Clusters may be grouped into different types, or motifs, which have specific algorithmic structure builders.
Here we consider motifs built from densely packed hard sphere models which form a seed, or atomic core for the metallic nanoparticles of interest.
Three dense packing configurations were used in this study; $N$ specifies the smallest building block for the atomic core:
\bei
    \item The cubic close-packed tetrahedron ($N=4$) yielding fcc clusters~\cite{kepler1611strena,HalesProofKeplerConjecture2005}
    \item The pentagonal bipyramid ($N=7$) which generate decahedral clusters~\cite{BagleyDensePackingHard1965}
    \item The icosahedron ($N=13$) used to build magic or Mackay icosahedra~\cite{Mackaydensenoncrystallographicpacking1962}
\ee
A diversity of different cluster geometries can be made by stacking layers of atoms in specific arrangements on top of the densely packed atomic seeds, and by truncating the growth along different high-symmetry directions~\cite{MartinShellsatoms1996}.
These structure-building algorithms, and others, are implemented in the Python atomic simulation environment \ase~\cite{Larsenatomicsimulationenvironment2017}.

The geometries which result from the different motif-specific truncation criteria can be classified as families, which share the same local atomic environment common to each motif, but differ in the topology of their polyhedral surfaces.
For example, in the \ase decahedron structure-builder, four parameters can uniquely specify a cluster model: a nearest neighbor bond distance, the number of layers parallel to the 5-fold axis, the number of layers truncated perpendicular to the 5 pentagonal edges, and the number of layers truncated perpendicular to the 5 apical vertices.
When no truncation exists, regular decahedra or pentagonal bipyramids are generated, whereas truncation of the pentagonal edges produces families of Ino-truncated wire-like decahedra~\cite{InoEpitaxialGrowthMetals1966} and apical truncation yields Marks decahedra~\cite{MarksExperimentalstudiessmall1994} with reentrant facets.
Changing the type and degree of truncation influences the resulting morphology of the cluster, and in decahedra, this also changes the relative number of atoms within the 5 fcc-like subunits versus the atoms situated at twin boundaries between the decahedral domains and at surfaces.

If a unique set of parameters that specify a cluster model is given as input to a structure-builder in \ase, a list of cartesian coordinates is returned which may be read in to a PDF calculating program.  In our case we use our own complex modeling infrastructure program, \cmi~\cite{juhas;aca15}.
PDFs are then calculated from the atomic coordinates using the Debye scattering equation~\cite{debye;ap15} PDF calculator implemented in \diffpy 's \dpc class under \srfit.
The atomic coordinates in space are held constant in the refinements but four parameters are allowed to vary to obtain good agreement between the calculated and measured PDFs: an isotropic expansion coefficient (linear scaling in $r$) to account for differences in nearest neighbor distances, a single U$_{iso}$ (isotropic ADPs), a single scale factor, and $\delta_2$, a parameter for correlated motion effects~\cite{proff;jac99}.
Parameters that describe the resolution of the measurement, (Q$_{damp}$ and Q$_{broad}$), are ideally obtained by independently refining a bulk calibrant measured in the same geometry as the nanocrystalline sample and fixed.

The cluster mine is built by iterating through the integer values for parameters, and combinations thereof, specifying the number of added and truncated layers for each motif-specific structure-builder.
The size of the structure mine (the number of clusters in the mine) can be tuned by providing bounds on the values that a given builder parameter may take, or by specifying a minimum and maximum number of atoms ($N_a$) in the clusters regardless of the builder.
During this procedure, \cmining stores metadata such as the number of atoms, atom type, nearest neighbor distance and motif, starting values for the refinable variables, along with the set of integers for that cluster.
This information is then passed to \ase which generates xyz coordinates of atoms, which are passed to \cmi that calculates the PDF and then refines the variable parameters against the measured PDF,  for each cluster in the mine.
The fit range in $r$ can also be adjusted prior to refining the library of clusters.
The \cmining program then returns a table of initial and refined PDF parameter values, and goodness of fit (R$_w$), with each individual refinement linked to the input cluster parameters and associated metadata.
A plot can then be generated of the best fit \rw\ vs. the number of atoms ($N_a$) for all clusters in the mine.  We call this plot the \minemap.
The \minemap can be filtered or labelled according to any cluster specific metadata, such as the motif.

The dimension of the input parameter space (typically 3-6) is significant and so the size of the mine can be large. For example, 2,419 unique combinations are possible for decahedra containing less than 1,500 atoms, including regular, Ino, Marks, and Ino-truncated Marks families.
However, the cluster-mining approach is easily parallelizable and  lends itself to deployment on multi-node computers.
This approach to nanostructure modeling may also be sped up by increasing the efficiency of selection of the clusters from the mine for testing, and we expect that statistical approaches such as machine learning will be effective in this regard, though this is beyond the scope of this paper.

\section{Results}

We first applied our cluster-mining approach to a PDF measured from sub-5~nm Pd nanoparticles that were described in~\cite{BanerjeeImprovedmodelsmetallic2018}.
In that work, the best cluster model that was found was a 609 atom regular decahedron with a maximum inter-vertex distance of 36.4~\AA.
This was determined by trial-and-error testing of a regular decahedral size series, starting with a 22.8~\AA\ (181 atom) decahedron, and ending with an 51.9~\AA\ (1442 atom) decahedron.
The refinement of the best-fit decahedral cluster core for the small Pd NPs is given in \npc(a) which shows the experimental nanoparticle PDF and the calculated PDF for the 609 atom decahedron, with the cluster structure reproduced in the inset.
The difference curves (fit residuals) for both the discrete cluster and fcc attenuated crystal (AC) models are offset below in blue and purple respectively.
\begin{figure}
\includegraphics[width=1\columnwidth]{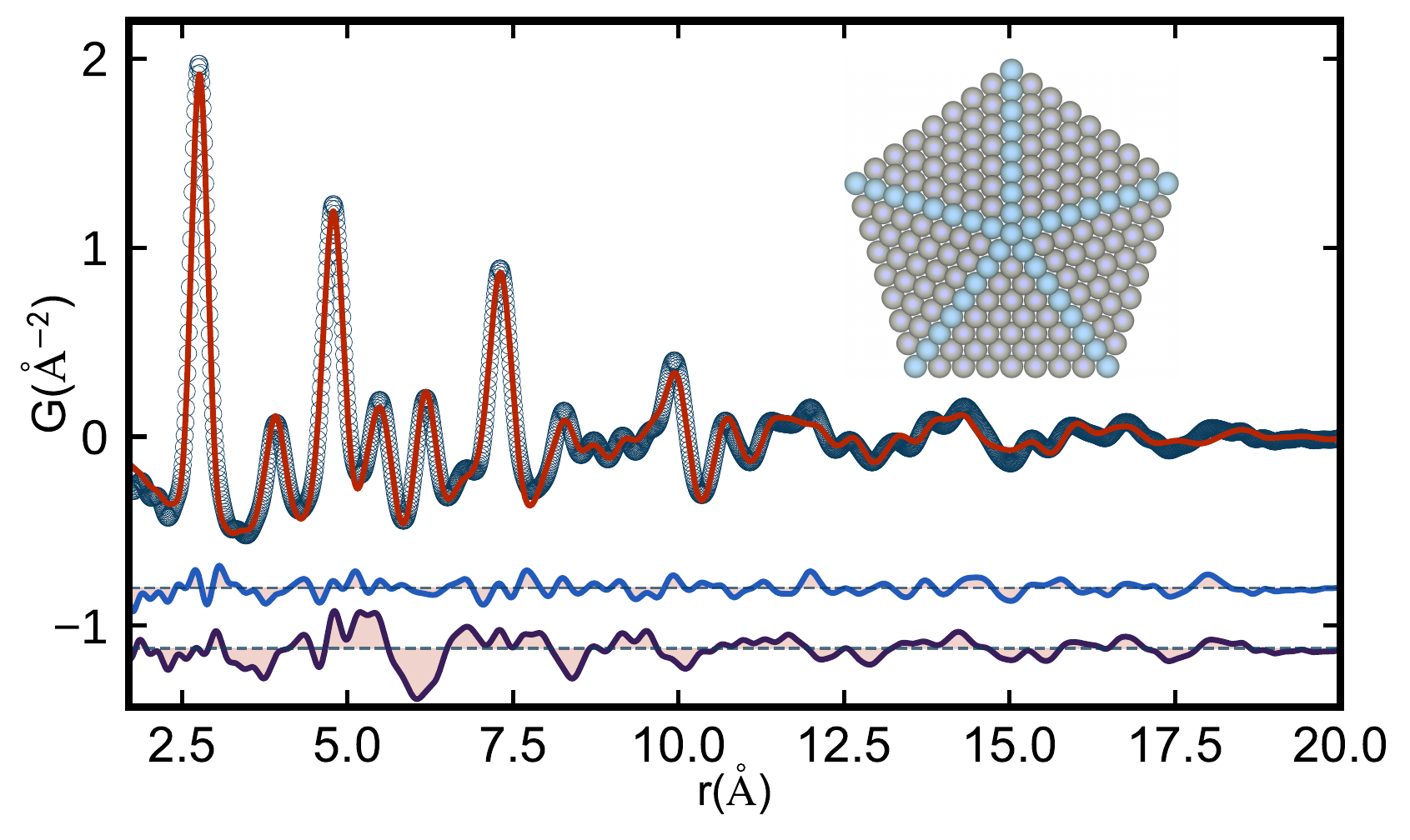}
\label{fig:npCore}
\caption{Experimental PDF (open circles) from \est3~nm Pd nanoparticles and the calculated PDF (red solid line) from a 3.6~nm decahedron (inset). Offset below are the difference curves from the discrete decahedral (blue) and spherically attenuated (AC) fcc crystal model (purple) refined to the measured Pd NP data.}
\end{figure}

In~\cite{BanerjeeImprovedmodelsmetallic2018} it was demonstrated that a diversity of small, representative clusters from motifs with different domain structures and morphologies were needed to fit all the metallic nanoparticle PDFs that were considered.
However, it is a laborious task to find the best cluster models, and it would also be valuable to know about the degeneracy of the solution set: how many different clusters give comparable agreement with the data.
To do this we can construct libraries, or mines, containing hundreds to thousands of discrete cluster models.
These were built combinatorially from motif-specific structure builders as described in Section~\ref{sec:modeling}.
To demonstrate what can be learned from this approach we applied it to the measured PDF from Pd nanoparticles shown in \npc\ by generating and fitting 464 different discrete models.
We start by investigating 60 clusters from a single structure motif (fcc) in greater detail.
The results are summarized in \pdfcc\ which shows the best-fit agreement factor of each fcc model plotted vs. the number of atoms in the model ($N_a$), which we call a \minemap.
We compare the cluster-mined solutions to that from the fcc AC model, which is the benchmark for refinements carried out in the traditional way using \pdfgui.  
For this Pd nanoparticle sample, the AC model resulted in an $R_w=0.253$, and this value is shown as a solid teal circle in \pdfcc.
This fit was obtained with a refined spherical particle diameter (SPD) of 19.4~\AA, which corresponds to $N_a\sim 225$ for a discrete FCC spherical cutout.
Next we built discrete spherical fcc cutouts to compare to the AC model.
These are shown as solid green circles with a dashed outline in \pdfcc.
This family of clusters has R$_w$'s that follow a trend with nanoparticle size.  The trend goes through a minimum at a particle size containing $N_a = 225$, the same as the AC model. 
Somewhat surprisingly, the R$_w$ of this model was lower than that of the AC model, though both correspond to spheres of fcc material. There are a number of differences between calculating the PDF of a spherical particle using a discrete spherical cluster and the Debye scattering equation (DSE) vs. a bulk model attenuated with the characteristic function of a sphere.  
One of the largest factors to affect the R$_w$ appears to be the choice of $Q_{min}$ used in the DSE calculation.  
This strongly influences the baseline in the PDF~\cite{farro;aca09} depending on the degree to which the small angle scattering signal is incorporated into the measured and calculated PDFs.  
Understanding this effect in detail is beyond the scope of this paper, but tests on this Pd nanoparticle sample showed that the best R$_w$ factors were obtained when the same $Q_{min}$ was used for the DSE calculations as was used in the treatment of the measured data.
We note that this careful study of spherical nanoparticle models yields insight into how the different cluster structures work with the data, and improvements in fit are possible over the AC model, but as was pointed out in~\cite{BanerjeeImprovedmodelsmetallic2018}, the spherical models do not remove much of the signal from residuals and are still deficient in many regards.

We now turn to models with the same fcc atomic structure, but which are cut out from the bulk with well defined surface faceting.  The clusters considered here were made by forming octahedral shapes exhibiting $\{001\}$  and $\{111\}$ facets.
Three families of faceted fcc octahedra are shown in \pdfcc: regular octahedra (solid diamonds) with only $\{111\}$ facets exposed, truncated octahedra (hexagons) with a mixture of $\{111\}$  and $\{001\}$ surfaces, and cuboctahedra (solid hexagons) which satisfy a specific truncation condition where the percentage of the surface covered by $\{001\}$ (non close-packed) facets is largest and all facet edges contain the same number of atoms.
The cuboctahedral family of clusters has the most isotropic or ``spherical" shape from the octahedral motif.
There are subtle variations in the \rw\ trends for each of the faceted fcc octahedral families with the cuboctahedral series following most closely the results of the discrete fcc spheres.
Regular and truncated octahedra follow trends that are offset slightly below the spherical and cuboctahedral series.
Overall, the fcc cluster families track very closely with each other, reaching \rw\ minima in the vicinity of $N_a\sim 250$ and in fact, the best candidate faceted octahedron is a slightly truncated cluster with 225 atoms, which has the same $N_a$ as the best fit discrete fcc sphere and AC approximation.
In the inset of \pdfcc\ we compare the residuals between the fcc AC model and the (a) minimum \rw\ fcc sphere and, (b) faceted octahedron, respectively.
Although small improvements are seen in $R_W$, it is clear that the majority of the misfit signal in the residual is not affected. This suggests that collectively, monocrystalline fcc cluster cores regardless of shape, might not be the most suitable structure motif for the small Pd NPs studied here.
\begin{figure}
\includegraphics[width=1\columnwidth]{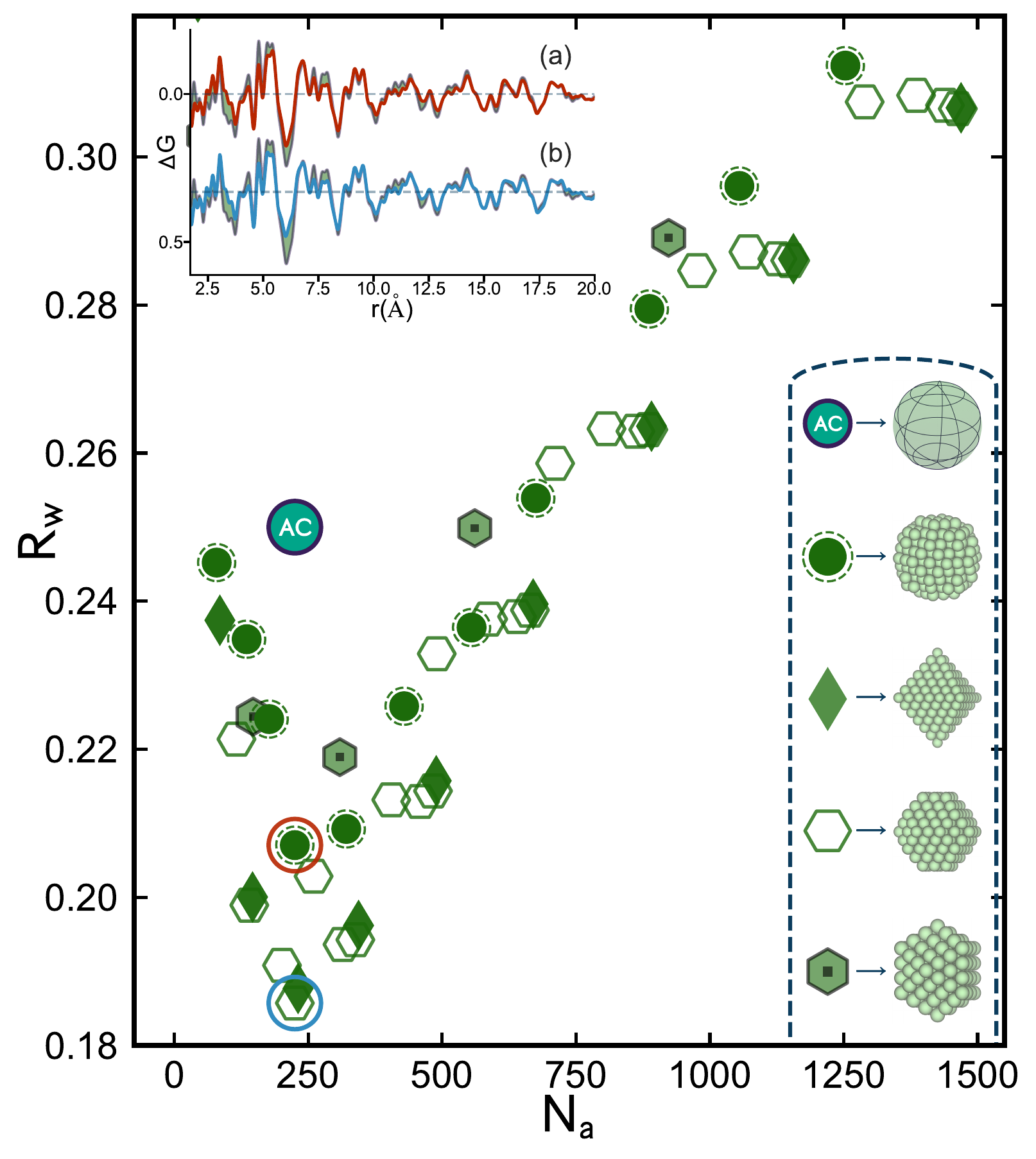}
\label{fig:PdFCC}
\caption{Scatter plot of agreement factors (\rw) for discrete fcc clusters fit to the Pd nanoparticle PDF, plotted as a function of the number of atoms per model ($N_a$). Each point is an individual PDF refinement of a discrete structure from a different fcc cluster type. These have been categorized as different families (see Section~\ref{sec:modeling} for details) which are represented in the legend at the bottom right. From top to bottom, the five families from the fcc motif shown here are AC, discrete spheres, regular octahedral, truncated octahedral, and cuboctahedral. In the scatter plot, the AC model fit is marked as a solid blue circle, and the best fit model from the discrete spherical and truncated octahedral families are highlighted with red and blue circles respectively. In the inset to the top left, the PDF fit residual from the AC model (light purple) is overlaid with the difference curves from the aforementioned best fit discrete sphere (a), and octahedral clusters (b), using the same colors as highlighted in the scatter plot.}
\end{figure}
Next, twinned cluster models from decahedral and icosahedral structure motifs were constructed and added to the mine and compared to the Pd nanoparticle data.
In \pdtwin\ we reproduce the same \rw\ scatterplot as discussed for fcc cutouts in \pdfcc\ with each point appearing as green symbols.
\begin{figure}
\includegraphics[width=1\columnwidth]{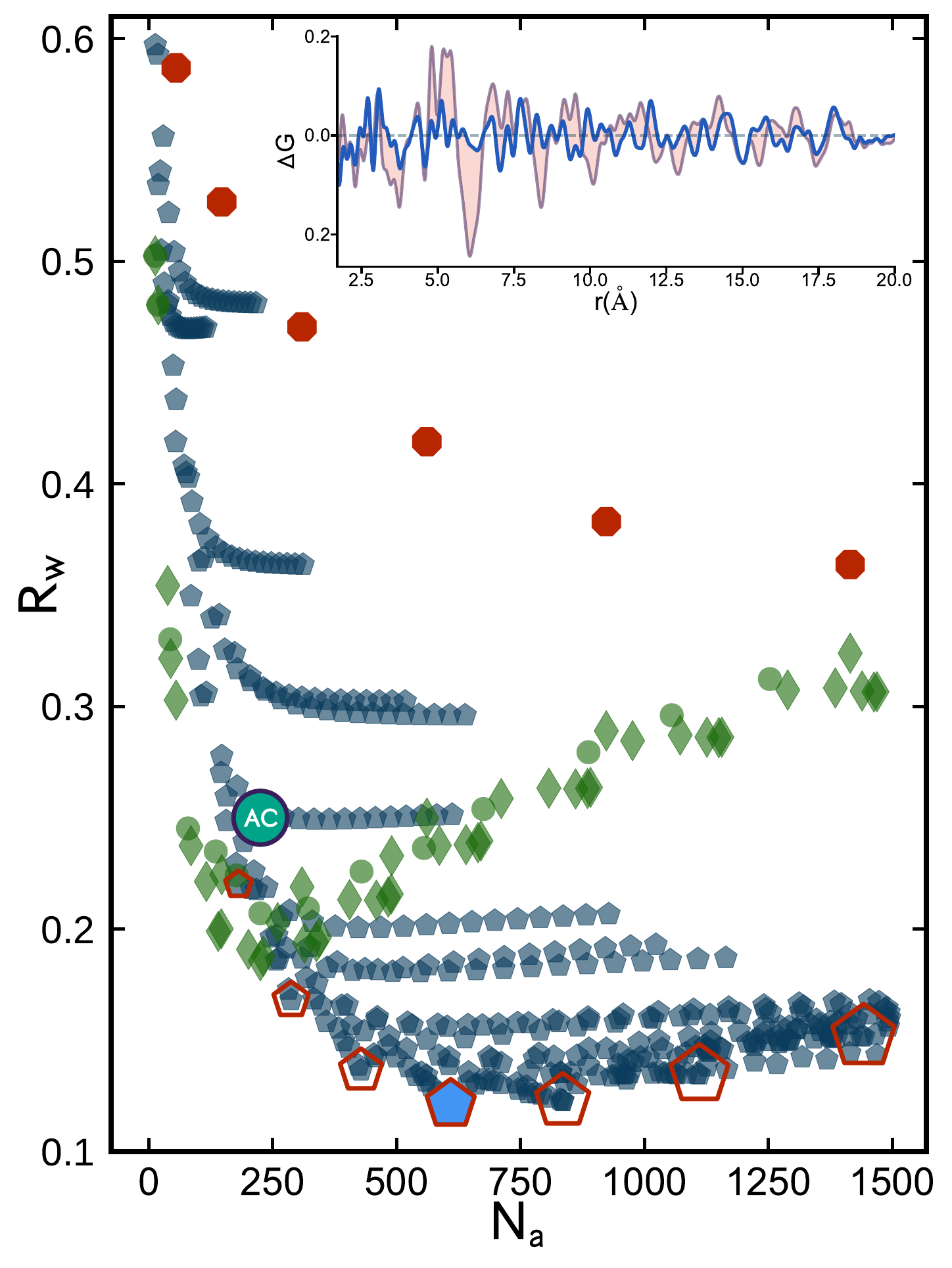}
\label{fig:PdTwin}
\caption{Scatter plot of agreement factors (\rw) for discrete clusters from three different structure motifs fit to the Pd NP PDF, plotted as a function of the number of atoms per model ($N_a$). Green diamonds and circles are for the fcc motif and include the faceted and spherical cluster families shown in \pdfcc. Red octagons are for Mackay icosahedra and blue pentagons are for different decahedral families (see text for details). The best fit AC model is marked as a solid blue circle. Red pentagons outline a size series of regular decahedra (pentagonal bipyramids). In the inset, the PDF fit residual from the AC model (light purple) is overlaid with the difference curve from absolute best-fit cluster model, which in this case is the 609 atom non-truncated decahedron (Inset~\npc).}
\end{figure}

The blue symbols are from 398 different decahedral structures including regular decahedra (pentagonal bipyramids), Ino decahedra, Marks decahedra, and Ino-truncated Marks decahedra (see the Methods section for additional details).  
The red symbols are from icosahedral structures.
55\% of the decahedral models tested are in better agreement with the measured Pd nanoparticle PDF than the best fit faceted fcc octahedron.
This can be seen as many of the blue symbols are at lower \rw\ values than the lowest green symbol in the \minemap.

\onecolumn
\begin{figure}
\includegraphics[width=1\columnwidth]{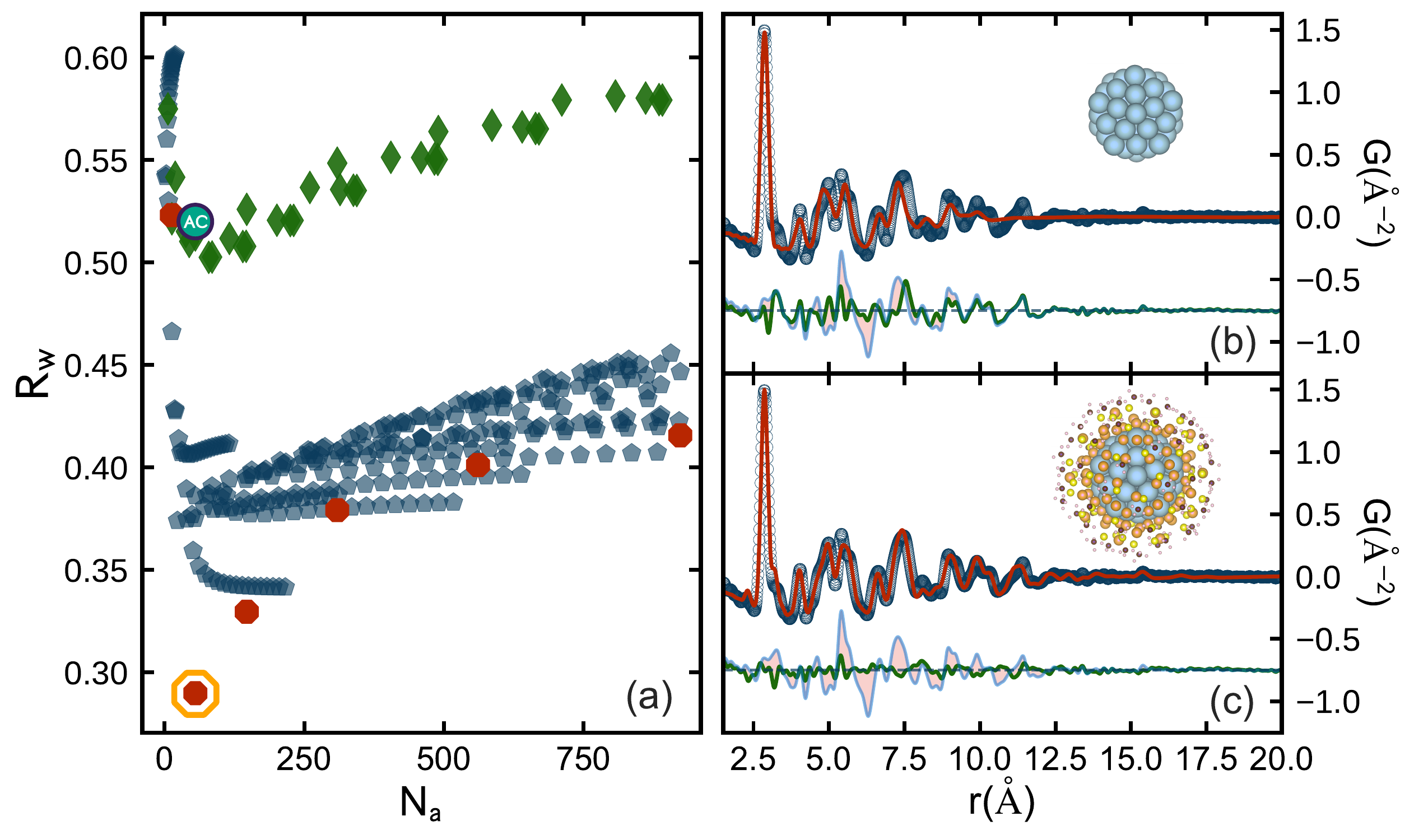}
\label{fig:SC6}
\caption{(a) Cluster-screen map for Au$_{144}$(SC6)$_{60}$ including structures from AC (teal), fcc octahedral (green), decahedral (blue) and icosahedral (red) motifs. The best-fit cluster core, a 55 atom Mackay icosahedron is outlined in orange (b) Measured PDF (open circles) from the Au$_{144}$(SC6)$_{60}$ cluster sample and the calculated PDF (red solid line) from the cluster-mined 55 atom Mackay core (shown in inset). The difference curve from this refinement is offset below in green, and overlaid with the AC residual in light blue. (c) Analogous to (b), except the calculated PDF (red solid line) is from a DFT derived structure solution~\cite{Lopez-AcevedoStructureBondingUbiquitous2009} for Au$_{144}$(SC6)$_{60}$ which shares the icosahedral core shown in (a), and also contains lower symmetry outerlayers. In the inset, the radii of atoms surrounding the DFT determined core are scaled down by a factor of 2 for illustration.}
\end{figure}
\twocolumn
The best candidate decahedral models for the Pd NP data turn out to be from a family of pentagonal bipyramids.
The \rw\ points from this family are outlined with red pentagons in \pdtwin.
These clusters increase in diameter, or maximum intervertex distance, as a function of $N_a$ and reach a minimum R$_w=0.121$ for a decahedron with 609 atoms and diameter of \est3.6~nm, which is nearly twice the size of the best fcc model and contains \est270\% more atoms.
This diameter for the 609 atom decahedron is much closer to the TEM estimated particle size of 3.0~$\pm$~0.3~nm for the Pd NPs investigated here.
The TEM estimate is not a full sample average, and is a slight underestimate of the average particle size.  This may be because the TEM estimate is averaging over particles viewed from different directions, and the particles are somewhat oblate in shape.
The shape of this 609 atom decahedron (\npc\ inset) also aligns with the observation of oblate-like morphologies in HRTEM images of these Pd NPs~\cite{BanerjeeImprovedmodelsmetallic2018}.
Most convincingly, in comparing the fit residual from the fcc AC model and the best fit decahedron (\pdtwin\ inset) we observe drastic changes to the largest amplitude features in the difference curve, with many of the misfit correlations removed altogether, which strongly supports the idea that the decahedral cluster core is capturing the correct modification to the fcc structure.

It is often discussed in the literature that the range of $r$ where features are seen in the PDF corresponds to a ``range of structural coherence" or a ``crystallite size" but this modeling shows how such a situation may come about.  
The observed PDF structural coherence range is actually roughly the size of one of the five fcc sub-domains that make up the decahedral cluster.  
This is an exemplar case where a model of a much larger cluster, which accounts for the inter-domain structure and domain twin boundaries, produces a significantly better fit to the PDF than just a model of incoherent small grains of fcc material and provides an illustration of how rather small nanoclusters may consist of sub-domains in general.
The other motif tested in \pdtwin, magic icosahedra (red markers), yield \rw's that are significantly worse than both the fcc and decahedral motifs, which shows that despite containing a high density of contact twin boundaries, the spatial arrangement of these domains is not representative for this Pd NP sample, and the icosahedral motif can be easily ruled out.

We now apply cluster-mining to a series of ultra-stable magic sized Au$_{144}$(SR)$_{60}$ clusters~\cite{WhettenNanocrystalgoldmolecules1996a} prepared with different thiolate ligands~\cite{AckersonSynthesisBioconjugationnmdiameter2010, QianAmbientSynthesisAu1442011}.
In \scsix(a) we show the \minemap\
from one sample in this series consisting of hexanethiol ligated clusters, Au$_{144}$(SC6)$_{60}$.
In this case, icosahedral structures perform better than the AC, fcc octahedral, and decahedral motifs.
The best fit model obtained is a 55 atom Mackay icosahedron with $R_w = 0.228$, highlighted with an orange outline in the \minemap (\scsix(a)).
In \scsix(b) we show the PDF of the best-fit cluster-mined 55 atom core. The difference curve is offset below and overlaid on the difference curve from the fcc AC approximation.
The main misfit in the AC difference curve between 5 and 8~\AA\ is drastically improved, and no other clusters are close in agreement, giving us confidence that the core of this Au$_{144}$ cluster is icosahedral in nature.
\begin{figure}
\includegraphics[width=1\columnwidth]{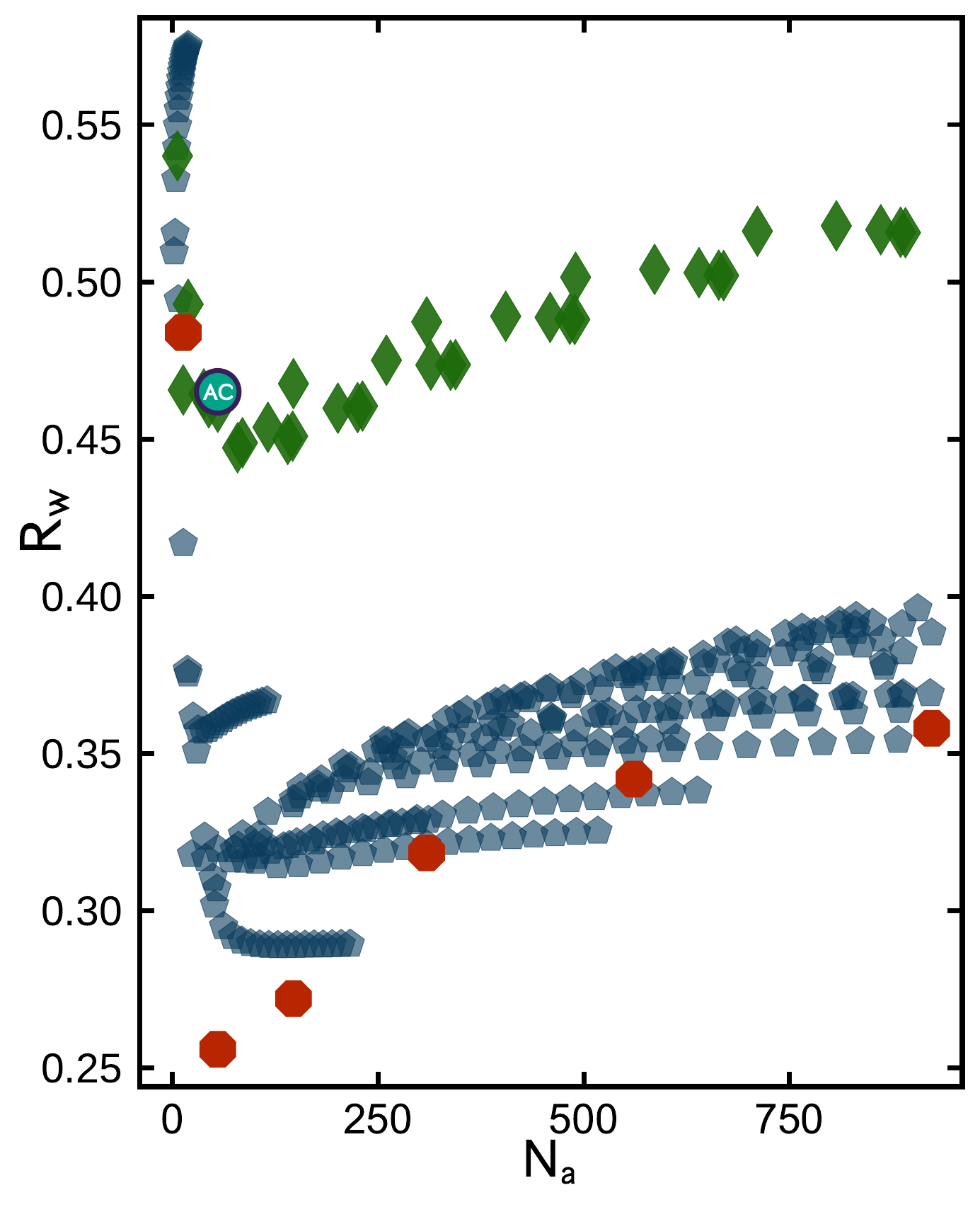}
\label{fig:polymorphism}
\caption{Cluster-screen map for a multi-phase cluster sample, Au$_{144}$(SC12)$_{60}$. The cluster-mine includes AC (teal), fcc octahedral (green), decahedral (blue) and icosahedral (red) motifs.}
\end{figure}
In this case, a structure solution for Au$_{144}$(SC6)$_{60}$ has been found by DFT, HAADF-STEM and PDF~\cite{Lopez-AcevedoStructureBondingUbiquitous2009,
BahenaSTEMElectronDiffraction2013,
jense;nc16}. 
In \scsix(c) we show the PDF from the 144 atom Lopez-Acevedo (LA) model, which contains chiral arrangements of atoms on top of a core that is nearly identical to a Mackay icosahedron~\cite{jense;nc16,BanerjeeImprovedmodelsmetallic2018}.
The additional lower symmetry outer layers of the LA model further remedies the misfit features at higher-$r$ (\scsix(c)) and improves the overall agreement factor to a value of R$_w=0.146$. 
This highlights the fact that cluster-mining can also identify good candidate cluster cores, which can be used as starting structures for more complex core/shell models.

Not all samples are ideally single phase, and we would like to know how robust the cluster-mining approach is in the case where more than one phase exists in the sample.
This can be tested using a Au$_{144}$(SR)$_{60}$ sample where a different thiolate ligand, dodecanethiol (SC12), was used to prepare the clusters.
This sample was shown to consist of both icosahedral and decahedral cores with the decahedral phase fraction being \est14\%~\cite{jense;nc16}.
The resulting \minemap is shown in \poly.
The cluster-mining methodology is stable, resulting in a \minemap that is largely similar to the pure, single-phase icosahedral SC6 sample shown in \scsix(a).  
It yields the 55 atom Mackay core as the best candidate cluster which is consistent with the expected majority phase, but the \minemap also shows that the R$_w$ trends for icosahedral and decahedral clusters has changed, with the two motifs reaching minima much closer to one another compared to the single-phase case. 
This behavior may be characteristic of nanoparticle mixtures.  
In the future we will explore extending cluster-mining to quantify minority phases in multi-phase samples.

\section{Experimental Methods}

Pd samples were prepared by the Murray group using methods described by~\cite{MazumderFacileSynthesisMPd2012}. 
Synthesis of Au$_{144}$(SR)$_{60}$ cluster samples was done in the Ackerson group following~\cite{QianAmbientSynthesisAu1442011}.
Pd nanoparticle data were collected at the National Synchrotron Light Source II (beamline XPD, 28-ID-2) at Brookhaven National Laboratory and data for the two cluster samples,  Au$_{144}$(SC6)$_{60}$ and  Au$_{144}$(SC12)$_{60}$ were collected at the Advanced Photon Source (11-ID-B), Argonne National Laboratory.
During both beamtimes, data were collected using the rapid acquisition PDF geometry~\cite{chupa;jac03} with large-area 2D detectors mounted behind nanopowder samples loaded in, or deposited on, polyimide capillaries and films. Pd NP samples were measured at 300~K with $\lambda=0.1846$~\AA\ and the two cluster samples were measured at 100~K with $\lambda=0.1430$~\AA.

\fittwod \cite{hamme;hpr96} was used to calibrate experimental geometries and azimuthally integrate diffraction intensities to 1D diffraction patterns for all three samples. Standardized corrections are then made to the data to obtain the total scattering structure function, $F(Q)$, which is then sine Fourier transformed to obtain the PDF, using \pdfgetxthree~\cite{juhas;jac13} within \xpdf~\cite{yang;arxiv15}.
The range of data used in the Fourier transform ($Q_{min}$ to $Q_{max}$, where $Q=4\pi\sin\theta/\lambda$ is the magnitude of the momentum transfer on scattering) was tuned per sample to give the best trade-off between statistical noise and real-space resolution, and also to truncate low-$Q$ scattering unambiguously originating from organic species in the sample.
For Pd NPs a range from $2.0 \leq Q \leq 26.0$~\AA$^{-1}$ was used, and for the cluster samples ranges of $ 0.8 \leq Q \leq 27.0$~\AA$^{-1}$ and $0.8 \leq Q\leq 26.0$~\AA$^{-1}$ were used for Au$_{144}$(SC6)$_{60}$ and  Au$_{144}$(SC12)$_{60}$, respectively.


\section{Acknowledgements}

Work  in the Billinge group was supported by the U.S. National Science Foundation through grant DMREF-1534910. 
Banerjee acknowledges support from the National Defense Science and Engineering Graduate Fellowship (DOD-NDSEG) program.
Juh\'{a}s was supported by the New York State BNL Big Data Science Capital Project under the U.S. Department of Energy contract No.~DE-SC-00112704.
Sample synthesis for the Pd nanoparticles discussed here was supported as part of the Catalysis Center for Energy Innovation, an Energy Frontier Research Center funded by the U.S. Department of Energy, Office of Science, Office of Basic Energy Sciences under Award Number DE-SC0001004.
Gold nanocluster data collected at the Advanced Photon Source at Argonne National Laboratory was supported by the U.S. Department of Energy, Office of Science, Office of Basic Energy Sciences (DOE-BES), under contract number DE-AC02-06CH11357.
Measurements of Pd nanoparticles were conducted on beamline 28-ID-2 (XPD) at the National Synchrotron Light Source II at Brookhaven National Laboratory, a DOE-BES user facility under contract No. DE-SC0012704.



\bibliographystyle{iucr}
\bibliography{billinge-group,abb-billinge-group,everyone,18sob_clusterMining}

\end{document}